\documentclass[aps,pra,twocolumn,showpacs,superscriptaddress,groupedaddress]{revtex4}  
\usepackage{graphicx}  
\usepackage{bm}        
\usepackage{amssymb}   
\usepackage{verbatim}   
\usepackage{gensymb}
\usepackage{subfigure}
\begin{document}
\title{Harmonic generation of noble-gas atoms in the Near-IR regime using
\textit{ab-initio} time-dependent R-matrix theory}
\author{O. Hassouneh} 
\email{ohassouneh01@qub.ac.uk}
\author{A.C. Brown}
\author{H.W. van der Hart}
\affiliation{Centre for Theoretical Atomic, Molecular and Optical Physics,
Queen's University Belfast, Belfast BT7 1NN,
UK}
\date{}

\begin{abstract}
We demonstrate the capability of {\em ab-initio} time-dependent R-matrix
theory to obtain accurate harmonic generation spectra of noble-gas atoms at
Near-IR wavelengths between 1200 and 1800
nm and peak intensities up to 1.8 $\times$ 10$^{14}$ W/cm$^2$. To accommodate the
excursion length of the ejected electron, we use an
angular-momentum expansion up to $L_{\rm max}=279$. The harmonic spectra show
evidence of atomic structure through the presence of a Cooper minimum in
harmonic generation for Kr, and of multielectron interaction through the giant resonance for Xe. The theoretical
spectra agree well with those obtained experimentally.
\end{abstract}

\maketitle

\section{Introduction}

Advances in laser technology have provided researchers with new
techniques to explore and exploit laser-matter interactions. 
Many experiments utilize the fundamental attosecond process-- high
harmonic generation (HHG)-- either as the source of high energy \cite{Pop12} and
ultrashort \cite{Pau01} light pulses or, more directly, as a window to
attosecond dynamics \cite{Bak07} or detailed molecular structure \cite{Vil04}.
These detailed spectroscopic techniques have shown that on ultrafast time-scales
both multiple ionization pathways \cite{Smi09} and multiple-electron
interference \cite{Shi11} are of fundamental importance in determining the
dynamics.

The main influence of atomic structure on HHG can be understood through the
so-called three-step model, in which a bound electron is first liberated by
tunneling ionization, then driven by the laser field, before recolliding with
its parent ion with the emission of a high-energy photon, all within a single
cycle of the driving field \cite{Cor93}. The recollision step can be regarded as
inverse photoionization, and therefore provides information on the basic
structure of the atom, while the time-energy mapping of the recolliding electron
gives a window to the attosecond-scale dynamics of the system. The highest
energy of the recolliding electron-- and thus, the extent of the
spectrum of emitted radiation-- is proportional to the square of the driving wavelength. Hence, in recent years,
HHG stimulated by Near-IR (NIR) radiation (wavelengths between 1200 and 1800 nm) has
gained interest. Although the total harmonic yield is significantly reduced at
these higher photon energies \cite{Tat07,Shi09}, the decrease in single-atom
efficiency can be compensated in the macroscopic medium by ensuring good
phase-matching \cite{Pop12}.

From a measurement point-of-view, the advantage of using long-wavelength
radiation is that the broad-bandwidth of the resulting HHG spectrum leads to
finer time-resolution. Furthermore, higher laser
intensities can be reached before the atom undergoes significant multiphoton
ionization. As
experiment continues to demonstrate the importance of multielectron effects in
ultrafast processes such as HHG, it is increasingly important that there is a
corresponding advance in theory, such that experimental data can be
accurately interpreted and interesting dynamics can be elucidated.

To the best of our knowledge, only the time-dependent configuration-interaction
singles (TDCIS) approach has been successfully applied to HHG in the NIR
regime for explicitly multielectron atoms \cite{Pab13,Pab14}.  In the present
report, we demonstrate the capability of the R-matrix including time-dependence
(RMT) approach to describe the harmonic response of general multielectron atoms
in the NIR regime from first principles. The approach has two defining
capabilities. Firstly, the code is optimized to run on massively parallel
($>1000$ cores) machines, thus making the extension to challenging
physical systems more amenable. Secondly, the RMT approach can be applied to
general atomic systems, including open shell atoms and ions, and can describe
atomic structure induced by electron-electron repulsion in detail.

Previously, we have applied time-dependent R-matrix (TDRM) theory to investigate
HHG at a wavelength of 390 nm \cite{Bro12,Bro13, Has14}. These studies
demonstrated that the details of the atomic structure can play a significant
role in HHG. The NIR radiation regime is substantially more challenging to
theory than the UV regime. For a wavelength of 1800 nm, an electron can absorb
many more photons from the field, leading to a substantial increase in the
angular momentum expansion. At an intensity of $1.8\times 10^{14}$W/cm$^2$ at
1800 nm the excursion length of a recolliding electron can extend beyond 200
$a_0$, so high-quality wavefunctions are required over extensive spatial
regions.  The implementation of time-dependent R-matrix theory in the TDRM codes
slows down significantly for these large expansions, making this approach
unfeasible. The RMT implementation performs much better for large
angular-momentum expansions \cite{Moo11}. Hence, this approach is more suitable
for the investigation of the atomic response to long-wavelength light fields.

To investigate the suitability of the RMT approach for calculations in
the NIR regime, we apply the approach to HHG
in Kr and Xe. These systems have been the subject of experimental
investigation with the Cooper minimum in Kr studied \cite{Shi12} as well as the effect of
the giant resonance in Xe \cite{Shi11}.  The giant resonance has already been
investigated using the TDCIS method \cite{Pab14}, but no comparison with experimental data was made.


\section{Time-dependent R-matrix Theory}

The RMT approach is the most recent implementation of time-dependent R-matrix
theory \cite{Moo11}. This approach adopts the standard R-matrix technique 
of separating the
physical system into two regions: an inner region, close to the nucleus, and an
outer region. In the inner region, all electrons interact strongly with
each other and the full Hamiltonian needs to be taken into account. In the outer
region, a single ejected electron is well separated from the other electrons, and 
exchange effects involving this ejected electron can be neglected.

A standard R-matrix basis is used to describe the wavefunction in the inner
region \cite{Bur11}. An $N$-electron atom is described as a direct product of
($N-1$)-electron states of the residual ion and a complete set of
single-electron functions representing the ejected electron. Additional
$N$-electron correlation functions can be added to the basis set to improve the
accuracy of the wavefunction.  
This is at variance with competitive methods such
as TDCIS, in which the wavefunction is expressed only in terms of the
Hartree-Fock ground state and singly excited configurations. At the
non-relativistic Hartree-Fock level, the RMT and TDCIS methods describe the
physics in the same manner, allowing for direct comparison of the methods. The
TDCIS method has the capability to account for the relativistic splittings of
residual ion states, giving access to spin-orbit dynamics in ultrafast
processes \cite{Pab14}. On the other hand, the RMT approach can include the
influence of double- and higher excitations in both the residual ion states, and
the full system under investigation. The two methods can therefore explore the
influence of a different set of interactions on the electron dynamics. Through
the inclusion of correlation orbitals, the RMT approach has the capability to
obtain an accurate description of the dynamics of open-shell systems as well as
closed-shell systems \cite{Mor14,Rey14,Bro12,Bro13}. The capability to account
for correlation in both the initial and residual ionic system is critical for
this accuracy.

In the outer region, the wavefunction is described in terms of
residual-ion states coupled with a finite-difference representation for the radial
wavefunction of the ejected electron \cite{Moo11,Lys11}.
The key distinction of RMT theory is the link between the
inner and outer regions. Standard R-matrix approaches connect different
regions through the so-called R-matrix \cite{Bur11,Lys09}. In the RMT approach, the
outer-region is connected to the inner region through the wavefunction itself. The outer-region
grid is extended into the inner region; the inner-region wavefunction
is evaluated on this grid and made available to the outer region. This suffices
to propagate the outer-region wavefunction. The inner region is connected to the outer region
in a traditional R-matrix fashion by determining a spatial derivative of the outer-region wavefunction
at the inner-region boundary. This derivative provides the boundary information needed to
update the inner-region wavefunction. 


The harmonic spectrum is determined by evaluating the time-dependent
expectation value of either the dipole moment or the dipole velocity \cite{Bro12b}. In
the present approach, the dipole acceleration is not suitable as
we investigate Kr and Xe, for which the inner orbitals must be kept frozen. The
harmonic spectra obtained through either the dipole moment or the dipole velocity show
the same spectrum up to well into the cut-off regime. They
differ by about 20\% in overall magnitude.
The main reason for this is the limited
ionic basis used in the calculations.

The use of highly scalable finite-difference techniques
  enables efficient parallelisation of the codes beyond 1000 cores.
This is key in the description of electron motion in long
wavelength driving pulses, as wavepackets can be driven very far from the
nucleus, necessitating a massively expanded outer region. For the investigation
of HHG, however, it should be noted that details of the outgoing wavepacket are
not as important when the electron is too far from the nucleus to return,
especially when HHG is determined using the dipole velocity operator. 

The noble-gas atoms, Kr and Xe, are described in R-matrix theory as a
direct product of residual-ion states (Kr$^+$ and Xe$^+$) coupled with a complete
set of outer-electron functions. In the present calculations, the residual-ion
states are described using Hartree-Fock orbitals for the ground-state of Kr$^+$
and Xe$^+$, respectively. We use these orbitals to generate $4s^24p^5$ and
$4s4p^6$ residual-ion states for Kr$^+$, and $4d^{-1}$, $5s^{-1}$ and $5p^{-1}$
initial states for Xe$^+$. To investigate the influence of the various shells
on HHG in Xe, we have carried out additional calculations in
which the $4d^{-1}$ residual-ion state is included or excluded. We have also
performed calculations including the $4p^{-1}$ state and find that its inclusion
does not affect the HHG spectra.

\section{Calculation Parameters}
\begin{figure}[t]
\includegraphics[width=7.8cm, angle=0]{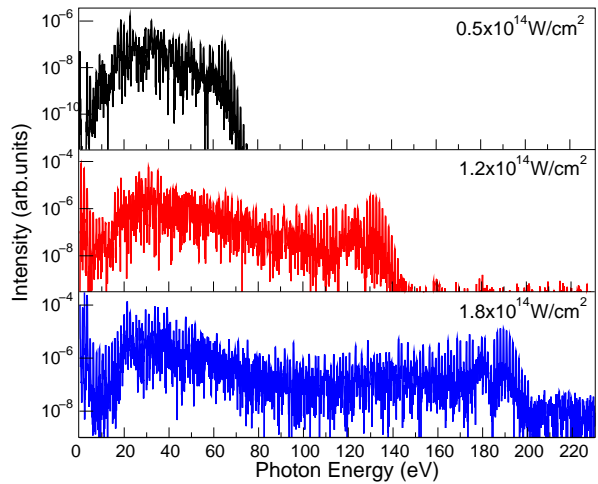}
\caption{(Color Online) Calculated harmonic spectra of Kr at a wavelength of 1800 nm 
for different peak laser intensities.}
\label{fig:kr}
\end{figure}

The description of Kr and Xe includes all available channels up to a maximum total
angular momentum $L_{\rm max}$ = 180 for Kr and 279 for Xe. The spectra were tested
for convergence with respect to angular momentum.  The time-step in
the wave function propagation for this calculation is normally set to 0.24 as.
Additional calculations were carried out at a shorter time-step of 0.12 as, with no
significant change in the spectra. The pulse
profile is given by a four-cycle sin$^{2}$ turn-on followed by two cycles at peak
pulse intensity and a four-cycle  sin$^{2}$ turn-off (4-2-4). To investigate the harmonic
spectrum for short pulses, additional spectra
were obtained for a pulse with a two-cycle sin$^{2}$ turn-on followed immediately
by a two-cycle sin$^{2}$ turn-off (2-0-2).

The R-matrix inner region has a radius of 20 $a_0$.  The continuum functions are described using a set
of 60 B-splines of order $k=9$, for each available angular momentum of the outgoing electron.
In the outer region we set the outer boundary to about 5000 $a_0$ to reduce
unphysical reflections of the wave function from this boundary. Re-scattered electrons
can have an energy of 10 $U_p$, with $U_p$ the ponderomotive potential \cite{tenup}.
This energy is about 550 eV in the present calculations. Hence, large box sizes
are required. The calculations were performed on ARCHER, the UK's supercomputing
facility, and typically employed around 1500 cores for 8 hours. 

\section{Results}

The main outcomes of the calculations are the time-dependent expectation values
of the dipole operator and the dipole velocity operator. These time-dependent
expectation values are Fourier transformed and squared to obtain the harmonic
spectrum.  Figure \ref{fig:kr} shows the harmonic spectrum obtained for Kr
irradiated by a 4-2-4 laser pulse, with a total duration of 60 fs, at a
wavelength of 1800 nm as a function of peak intensity.  The spectra show the
standard form of odd harmonics of the fundamental photon energy with a
well-defined plateau region and a cut-off energy of 70 eV at 0.5 $\times$
10$^{14}$ W/cm$^2$ increasing to 188 eV at 1.8 $\times$ 10$^{14}$ W/cm$^2$.
This is in line with the cut-off formula: $E_c = 1.3 I_p +3.17U_p$, where $I_p$
is the ionization potential and $U_p$ the ponderomotive potential \cite{Lew94}. 


\begin{figure}[t]
\includegraphics[width=7.8cm,angle=0]{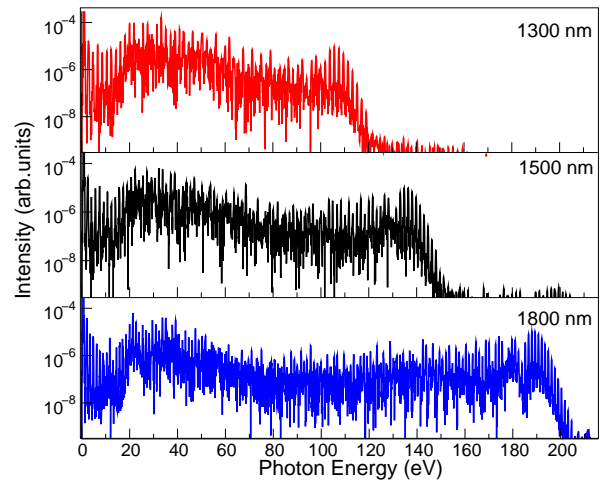}
\caption{(Color online) Calculated harmonic spectrum of Kr at a peak intensity of 1.8$ \times$ 10$^{14}$
W/cm$^2$ for different laser
wavelengths.}
\label{fig:lambda}
\end{figure}

At the highest peak intensity of 1.8$\times$10$^{14}$ W/cm$^2$, Fig. \ref{fig:kr}
shows a clear influence of atomic structure on the harmonic spectrum through a
minimum in the harmonic yield at a photon energy of around 85 eV. This minimum is also
visible at a peak intensity of 1.2$\times$10$^{14}$ W/cm$^2$, but at 
0.5$\times$10$^{14}$ W/cm$^2$, the harmonic plateau does not extend sufficiently
far.  This is a so-called Cooper minimum, in which the radial dipole matrix
element between the $4p$ orbital in Kr and the continuum $d$ orbital vanishes.
The photorecombination step in the three-step model is the reverse process of
photoionization, and it will thus be affected similarly: when the radial matrix
element vanishes, photorecombination is not allowed. This minimum has been
observed in photoionization spectra and occurs at a photon energy around 84 eV
\cite{Sam02}, in line with the present observations. 

Figure \ref{fig:lambda} 
shows the dependence
of the harmonic cut-off on laser wavelength for Kr. We observe a cut-off energy of 115 eV at 1300
nm, 140 eV at 1500 nm and 188 eV at 1800 nm, in line with the cut-off formula
\cite{Lew94}. The position of the minimum in the harmonic yields is largely
independent of driving wavelength. At the shortest wavelength, the spectrum
barely extends to the energy of the Cooper minimum. Unambiguous observation of
the minimum is only possible at the longer wavelengths.

\begin{figure}[b]
\includegraphics[width=7.8cm, angle=0]{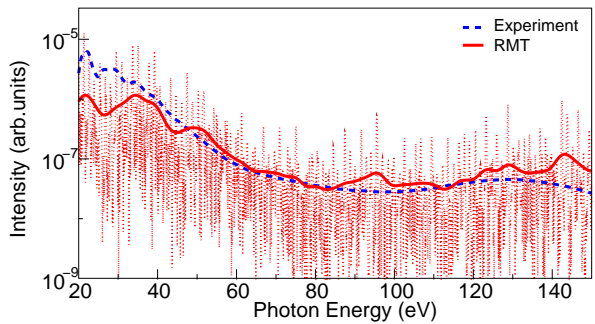}
\caption{(Color Online) HHG spectrum for Kr calculated using RMT (dotted, red
  line), smoothed spectrum (solid, red line) and the experimental spectrum
  (dashed, blue line, \cite{Shi11}) at a wavelength of 1800 nm and
a peak intensity of 1.8
$\times$ 10$^{14}$ W/cm$^2$. Although the theoretical spectrum extends to 190 eV, the
experimental spectrum was only reported for photon energies up to 150 eV. The
experimental data has been renormalised for comparison with the theoretical spectrum.}
\label{fig:krexp}
\end{figure}

Figure \ref{fig:krexp} shows a comparison with the experimental HHG spectrum
\cite{Shi12}. Although the theoretical spectrum extends up to 190 eV,
the experimental spectrum was presented only for photon energies up to 150 eV. Figure
\ref{fig:krexp} shows excellent agreement between the two spectra, demonstrating that
RMT theory has the capability to predict harmonic spectra in the NIR regime.
We note, however, that while RMT calculates single-atom HHG spectra, the experimental spectra include
propagation effects, which may indicate a preference for either the so-called
short or
long trajectories \cite{Smi13}. 

Figure \ref{fig:krexp} suggests the minimum appears between 90-100
eV in experiment, whereas the calculated minimum appears closer to 85 eV.
However, the position of the theoretical minimum is difficult to assign
unambiguously, as there is variation in intensity between neighbouring
harmonics. Furthermore, the presented experimental data is an averaged spectrum,
and the different rates of increase around the minimum may have shifted the
experimentally observed minimum to higher energy. 

\begin{figure}[t]
\includegraphics[width=7.8cm, angle=0]{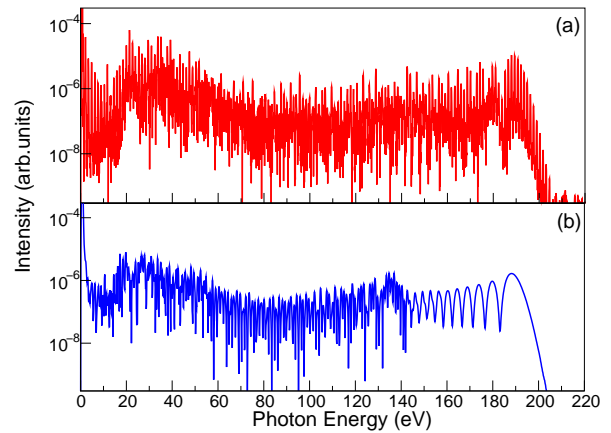}
\caption{(Color Online) The HHG spectrum of Kr at a wavelength of 1800 nm and a
  peak intensity of 1.8 $\times$ 10$^{14}$ W/cm$^2$ 
for a 10-cycle pulse
with a 4-2-4 profile (a) and a 4-cycle pulse with a 2-0-2
profile (b).}
\label{fig:pulse}
\end{figure}

The comparison between experiment and theory is more difficult for harmonic
photons with an energy greater than 150 eV. One of the reasons for this
difficulty is the experimental pulse length. In Fig. \ref{fig:pulse}, we
compare the harmonic spectrum obtained for our current 4-2-4 pulse with a pulse
consisting of a 2-cycle $\sin^2$ turn-on, followed by a 2-cycle $\sin^2$
turn-off. For both profiles, the carrier-envelope phase is set to $0\degree$
The full-width at half-maximum of the 2-0-2 pulse is about 8.7 fs, which
is comparable with the 10 fs pulse used in the experiment. This reduction
in pulse length has a significant effect on the harmonic spectrum. Classical
calculations show that only one electron trajectory leads to recollisions with
energies greater than 140 eV. This leads to
a broad spectrum as peaks separated by twice the
fundamental photon energy require interference between multiple trajectories. 

We have investigated the influence of the multielectron interaction on HHG in Xe for a
4-cycle, $1.9\times10$ W/cm$^2$, 1800 nm pulse.
The photoionization spectrum
of Xe is dominated by the so-called giant resonance, in which a $4d$
electron is excited into a short-lived quasi-bound resonance of $f$ character \cite{Xe}.
However, this resonance cannot be thought of as a single-electron effect: partial
photoionization spectra demonstrate evidence of the resonance in the photoemission
of the $5s$ and the $5p$ electron, as well as photoemission of the $4d$ electron.

The most likely electron to be removed from the Xe atom in the ionization step
of the recollision model is the outermost $5p$ electron, particularly at long
wavelengths. The observation of the giant resonance
is then a clear indication of electron-electron interactions: a $4d$ electron
interacts with the recolliding electron to be
promoted into the $5p$ hole. This allows the recolliding electron to recombine
into the $4d$ shell-- for which the photorecombination matrix elements are
particularly large relative to the $5p$ matrix elements-- leading to a broad
resonance in the HHG yield.
To demonstrate this latter point, calculations have been carried out which
either allow or disallow emission of one of the $4d$ electrons.

\begin{figure}[b]
\includegraphics[width=7.8cm]{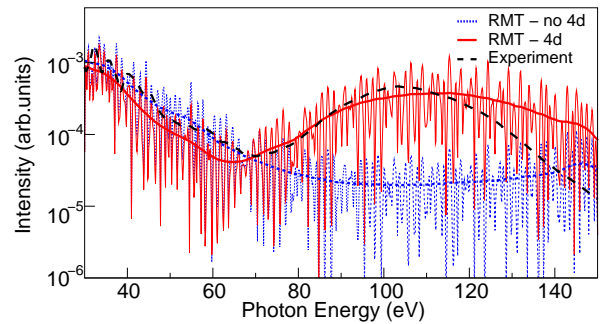}
\caption{(Color online) The harmonic spectrum for Xe irradiated by
an 11 fs, 1800 nm laser pulse with a peak
intensity of 1.9 $\times$ 10$^{14}$ W/cm$^2$ as obtained in experiment
(dashed, black line, \cite{Shi11})
and using the present approach, either including emission of a 4d electron
({\it solid}, red line) or
excluding this emission (dotted, blue line). The thick lines indicate the smoothed theoretical
results. The experimental data has been renormalised for comparison with the
theoretical spectra.}
\label{fig:xe}
\end{figure}
Figure \ref{fig:xe} shows the calculated harmonic spectrum for Xe irradiated
by 1800 nm laser light with a peak intensity of $1.9\times$10$^{14}$ W/cm$^2$.
The figure compares the harmonic spectrum obtained both by including and
neglecting the influence of
the $4d$ electron. In line with TDCIS calculations \cite{Pab13}, opening the $4d$ shell increases the harmonic yield by
about an order of magnitude in the neighbourhood of the giant resonance at
around 115 eV. The calculated spectrum shows good
agreement with experiment \cite{Shi11} up to a photon energy of about 100 eV, when the experimental harmonic
yield drops off faster than the theoretical ones. Thus the theoretical resonance
appears to be broader than the experimental resonance which
is centered on 100 eV. Again, these differences may be due to macroscopic
propagation effects. However, the description of the resonance might also be improved
by including more electron-correlation effects, to describe changes in the $4d$ and
$5p$ orbitals during HHG. Although it
is, in principle, possible to include additional orbitals to account for these changes, their
inclusion would significantly increase the number of residual-ion states to
account for, rendering the calculations unfeasible at present. 

\section{Conclusions}
In conclusion, we have demonstrated the feasibility to investigate HHG in
general atomic systems in the NIR regime from first principles with full inclusion of
electron-electron repulsion. 
Specifically, we have applied the RMT codes to investigate HHG in
Kr and Xe at wavelengths up to 1800 nm. Experimental spectra show
a significant influence of atomic structure, which is well reproduced in the theoretical
calculations. The
present investigations only concern noble-gas atoms, as these are the systems of main experimental
interest. However, the RMT approach has already been applied to photoionization of
atomic systems with open outer shells, such as Ne$^+$ and C \cite{Mor14,Rey14}. The approach
should therefore be capable of obtaining HHG spectra for general atomic systems
in the NIR regime.

HHG is only one of the many aspects of laser-matter interactions
that has been investigated experimentally. For example, experiment has also
investigated ejected-electron momentum distributions for atoms and ions
irradiated by short 1800 nm fields. The theoretical description of these
distributions is significantly more sensitive to the number of total angular
momenta included in the calculations. It will therefore be interesting to
explore whether other aspects of laser-matter interactions at 1800 nm can
also be investigated accurately using the RMT approach.

The authors thank David  Villeneuve and Andrew Shiner for providing the experimental data in numerical
form, and Stefan Pabst for additional, theoretical data.
OH acknowledges financial support from the University of Jordan. HWH acknowledges
financial support from the UK EPSRC under grant no. EP/G055416/1 and the EU Initial
Training Network CORINF. This work used the
ARCHER UK National Supercomputing Service (http://www.archer.ac.uk).

\bibliography{/users/abrown41/Documents/pub/HG-MidIR/mybib.bib}

\end{document}